**Submicron scale tissue multifractal anisotropy in polarized light scattering**

Nandan Kumar Das[1], Semanti Chakraborty[1], Rajib Dey[1], P. K. Panigrahi[1], Igor Meglinski[2,*], and Nirmalya Ghosh[1,*]

*Corresponding author: Igor.Meglinski@oulu.fi, nghosh@iiserkol.ac.in

[1] *Department of Physical Sciences, Indian Institute of Science Education and Research Kolkata Mohanpur-741246, West Bengal, India*

[2] *Opto-Electronic and Measurement Techniques, University of Oulu, P.O. Box 4500, Oulu, 90014, Finland*

**Abstract**

A number of disordered systems exhibit local anisotropy in the fractal or multifractal correlation and in the resulting scaling behavior, which contain wealth of information on the system. Here, we demonstrate that the spatial dielectric fluctuations in a random medium like biological tissue exhibit such multifractal anisotropy, leaving its unique signature in the wavelength variation of the light scattering Mueller matrix and manifesting as an intriguing spectral diattenuation effect. We have thus developed an *inverse* analysis method for the quantification of the multifractal anisotropy from the scattering Mueller matrix. The method is based on processing the relevant Mueller matrix elements in Fourier domain using Born approximation followed by multifractal analysis. Application of this technique on tissues of human cervix *ex vivo* demonstrate the potential of the multifractal anisotropy parameters as novel biomarkers for screening subtle micro-structural changes associated with precancers. Sensing structural anisotropy in the sub-micron length scale via the multifractal anisotropy parameters may prove valuable for non-invasive characterization of a wide class of complex materials and disordered scattering media.

**1. Introduction**

Self-similar structures and processes have been the focus of multi-disciplinary research due to their fundamental nature and applications in diverse branches of science [1–8]. Typically, monofractal measures, characterized by a single scaling exponent have been used to model most of the fractal systems [1]. However, a single scaling exponent may not often be adequate to describe many naturally occurring complex self-similar structures and processes [2–8]. More generalized treatments based on the so-called multifractal models, dealing with a spectrum of scaling exponents, have thus been developed [2,3]. In addition to multifractal behaviour, a few special class of disordered systems exhibit anisotropy in the scaling behaviour [9–12]. Directional anisotropy in the local microscopic fluctuations may eventually manifest as multifractal anisotropy [9,10]. For example, such anisotropy in has been observed in anisotropic turbulence and the corresponding correlation function has been modeled by using expansion in terms of the irreducible representation of the relevant rotational symmetry group [11,12]. Quantification of such multifractal anisotropy potentially yields wealth of information on the ultra-structure and disorderness of the system.

Measurements of wavelength or angular variation of elastically scattered light in combination with *inverse* analysis can be used to quantify self-similarity in the spatial distribution of refractive index (RI) of random scattering media (such as biological tissues) [13,14]. Such information on the fractal/multifractal nature of spatial RI fluctuations of tissues has shown promise for early detection of cancer [14,15]. For probing anisotropies in RI variations, one further needs to invoke measurements involving polarization. Mueller matrix (a 4×4 matrix), the transfer function of polarized light's interaction with medium, contains all the relevant information on anisotropy [16,17]. The two basic medium anisotropy properties are diattenuation (dichroism) and retardance (birefringence). The former represents amplitude anisotropy dealing with differential attenuation and the latter represents phase anisotropy (differential phase) between orthogonally polarized light [16,17]. Conventional Mueller matrix measurements are used to quantify *macroscopic* anisotropy for the characterization of tissue and other complex materials [16,17]. In this paper, we demonstrate that the spatial RI fluctuations of a random medium like tissue exhibits multifractal anisotropy. It is shown that the multifractal anisotropy leaves its unique signature in the wavelength variation of the scattering Mueller matrix elements, manifesting as an intriguing spectral diattenuation effect. We have therefore addressed the corresponding *inverse* problem of extracting information on the *microscopic* anisotropy in multifractal scaling of RI from the scattering Mueller matrix. The *inverse* analysis is based on processing the wavelength dependence of the relevant scattering Mueller matrix elements (encoding the diattenuation effect) in the Fourier domain using Born approximation followed by multifractal analysis. The model is applied on spectral Mueller matrices recorded from *ex-vivo* tissues of human cervix of different pathology grades. The newly defined *multifractal anisotropy* parameters appear sensitive in detecting subtle (otherwise hidden) changes in tissue ultra-structure in the sub-micron length scales, associated with precancerous morphology.

## 2. Theory

### 2.1 Scattering Mueller matrix modeling in Born approximation

Biological tissue can be modeled as a scattering medium having continuous random variations of RI, $n(r) = n_0[1 + \delta n(r)]$ [18–21]. Here, $r$ is the location within the volume and $n_0$ is the average RI; $\delta n(r)$ is the fractional RI fluctuations giving rise to scattering. The scattered electric field from such medium can be modeled by conventional 2×2 amplitude scattering matrix [22]. For an *anisotropic* weakly fluctuating (small $\delta n(r)$) scattering medium, the elements $s_2$ and $s_1$ of the scattering matrix, which represents, respectively, the scattered fields with the polarization parallel and perpendicular to the scattering plane, can be modeled by using the first order Born approximation [14,15,19,22]. The $s_2$ and $s_1$ elements can be obtained by representing the scattered field as a Fourier transformation of the scattering potential (determined by $\delta n(r)$) and by incorporating the anisotropic distribution of index fluctuations [15,19,22]:

$$s_2(\theta,\lambda) \approx \cos\theta \times \Lambda_{\|}^2 \int \eta_{\|}(r)\, e^{i(\boldsymbol{\beta}\cdot r)} d^3 r\,,$$

$$s_1(\theta,\lambda) \approx \Lambda_{\perp}^2 \int \eta_{\perp}(r)\, e^{i(\boldsymbol{\beta}\cdot r)} d^3 r\,. \tag{1}$$

Here, $\boldsymbol{\beta}$ is the scattering vector with modulus $\beta = 2k \sin\left(\frac{\theta}{2}\right)$, $\beta = 2\pi\nu$, $\nu$ the spatial frequency; $k = 2\pi/\lambda$ is the wave vector, $\lambda$ and $\theta$ are the wavelength and the scattering angle; $\Lambda_{\|/\perp} = \lceil n_0 \delta n \rceil_{\|/\perp}$ the index fluctuation strengths for orthogonal linear polarizations; $\eta_{\|/\perp}(r)$ are the corresponding (along the two orthogonal directions) index inhomogeneity distributions. On practical grounds, we have assumed that – (a) the index variations arise from statistical inhomogeneities having different spatial dimensions, albeit with similar amplitude of fluctuating index $\delta n$ [15,20] and (b) the index fluctuations exhibit *uniaxial* anisotropy, i.e., it can be decomposed into two orthogonal components, parallel and perpendicular to the axis of the anisotropy [16,17].

For such an *anisotropic* fluctuating medium with the axis of anisotropy oriented at an angle $\varphi$ with respect to the laboratory polarization axis, the 4×4 scattering Mueller matrix [$M(\theta,\lambda)$] describing the Stokes vector (**I**) transformation can be obtained as [17]:

$$I_0 = k^4[T^{-1}(\varphi)S(\theta,\lambda)T(\varphi)]I_i = M(\theta,\lambda)I_i \qquad (2)$$

Here, $T(\varphi)$ is the conventional 4×4 rotation matrix in the Stokes polarization space, and the elements of the matrix $S$ have standard relations with $s_2$ and $s_1$ [11,12,17].

Using Equations (1) and (2), $s_2$ and $s_1$ can be related to the elements of $M(\theta,\lambda)$ as:

$$|s_2(\theta,\lambda)|^2 = k^{-4}\left\{M_{11}(\theta,\lambda) + \sqrt{M_{12}^2(\theta,\lambda) + M_{13}^2(\theta,\lambda)}\right\} = \cos^2\theta \times \Lambda_{\|}^4 \left|\int \eta_{\|}(r)\, e^{i(\beta..r)} d^3r\right|^2,$$

$$|s_1(\theta,\lambda)|^2 = k^{-4}\left\{M_{11}(\theta,\lambda) - \sqrt{M_{12}^2(\theta,\lambda) + M_{13}^2(\theta,\lambda)}\right\} = \Lambda_{\perp}^4 \left|\int \eta_{\perp}(r)\, e^{i(\beta..r)} d^3r\right|^2. \qquad (3)$$

In the monofractal approximation, the right hand side of Equation (3) (the power spectrum of the spatial RI fluctuations) assumes a power law behavior [14]:

$$|s_2(\nu)|^2 \approx \nu^{-\gamma_{\|}}, \; |s_1(\nu)|^2 \approx \nu^{-\gamma_{\perp}} \qquad (4)$$

Equations (3) and (4) provide a recipe for the determination of fractal anisotropy (differential power law exponent for orthogonal linear polarizations, $\Delta\gamma = |\gamma_{\|} - \gamma_{\perp}|$) from the wavelength variation of experimental scattering Mueller matrix (recorded at fixed θ). However, the monofractal approximation may not be realistic in tissues, rather, the index fluctuations may exhibit multifractality [15,23]. We, therefore, adopt the following *inverse* analysis strategy for the quantification of multifractal anisotropy. Equation (4) implies that the multifractality in the spatial distributions $\eta_{||/\perp}(r)$ are encoded in the Fourier domain in $|s_2(v)|^2$ and $|s_1(v)|^2$. Consequently, the multifractality information can in principle be obtained from

$$\eta_{||/\perp}(\rho) \sim \int |s_{2/1}(\beta = 2\pi v)| e^{-i(\beta \cdot r)} d^3\beta. \tag{5}$$

Here, $\eta_{||/\perp}(\rho)$ may be interpreted as statistically equivalent index inhomogeneity distributions with length scale $\rho = |r - r'|$ (the distance between any two points in the medium), along two orthogonal directions. The parameters $\eta_{||/\perp}(\rho)$ can then be analyzed via multifractal detrended fluctuation analysis to quantify the multifractal anisotropy of RI fluctuations.

Note the elements $M_{12}$ and $M_{13}$ [in Equation (3)] represent linear diattenuation effect (differential scattering for orthogonal linear polarizations) with magnitude $d = \frac{\sqrt{M_{12}^2 + M_{13}^2}}{M_{11}}$ [17]. Apparently, the magnitude of $d$ should vanish for backscattering ($\theta = 180°$) from any *isotropic* weakly fluctuating scattering medium. In contrast, $d$ should be non-zero for a scattering medium exhibiting fractal (multifractal) *anisotropy*. Thus, the unique signature of multifractal anisotropy is encoded in the wavelength variation of $d(\lambda)$ (experimental evidence is presented subsequently).

*2.2 Multifractal Detrended Fluctuation Analysis (MFDFA)*

The details of the multifractal detrended fluctuation analysis (MFDFA) can be found elsewhere [2,23,24]. Briefly, the fluctuation profile $Y(i)$ (spatial series of length N, $i = 1 \ldots N$) is divided into $N_s = \text{int}(N/l)$ segments $m$ of equal length $l$. The local trends ($y_m(i)$), determined by least square polynomial fitting of each segment $m$, are subtracted from the profile to obtain the detrended fluctuations and the resulting variance as:

$$F^2(m,l) = \frac{1}{l}\sum_{i=1}^{l}[Y\{(m-1)l+i\} - y_m(i)]^2. \tag{6}$$

The moment ($q$) dependent fluctuation function is constructed by averaging over all segments

$$F_q(l) = \left\{\frac{1}{2N_s}\sum_{m=1}^{2N_s}[F^2(m,l)]^{\frac{q}{2}}\right\}^{1/q}. \tag{7}$$

The multifractal signal can be characterized by approximating a power law behavior of the fluctuation function $F_q(l) \sim l^{h(q)}$. Here, $h(q)$ is the generalized Hurst scaling exponent, which is related to the conventional classical multifractal scaling exponent $\tau(q)$ as [24-28]:

$$\tau(q) = qh(q) - 1. \tag{8}$$

Note that monofractal signal displays linear $\tau(q)$ spectra, $\tau(q) = qH - 1$, where $H$ is the global Hurst exponent ($H \in (0,1)$) [24]. Moreover, $\tau(q=2)$ is related to the power law exponent of the Fourier power spectrum as $\gamma = \tau(q=2) + 2$ [8]. For multifractal signal, on the other hand, $\tau(q)$ is a non-linear function of $q$. Multifractality is subsequently characterized via $\tau(q)$, and the singularity spectrum $f(\alpha)$. These are related as [24]

$$\alpha = \frac{d\tau}{dq}, \quad f(\alpha) = q\alpha - \tau(q), \tag{9}$$

where $\alpha$ is the singularity strength or Holder exponent. The quantity $\sigma$, the full width of $f(\alpha)$ (defined as the difference in the minimum and the maximum value of $\alpha$ corresponding to the two minima of $f(\alpha)$) [7,8] measures the strength of multifractality.

In what follows, we (i) provide experimental evidence of multifractal anisotropy in spatial variation of tissue RI by *forward* analyzing differential interference contrast (DIC) images via the MFDFA; then (ii) analyze, derived from the spectral Mueller matrix, $\eta_{||}(\rho)$ and $\eta_{\perp}(\rho)$ by

MFDFA to quantify the multifractal anisotropy parameters of tissue - differential classical scaling exponent, $\Delta\tau = |\tau(q = 2)_{||} - \tau(q = 2)_{\perp}|$, and differential width of singularity spectrum $\Delta\sigma = |\sigma_{||} - \sigma_{\perp}|$ for orthogonal linear polarizations.

## 3. Experimental Section

The experimental system for backscattering spectroscopic Mueller matrix measurement (**Figure 1**) comprises a Xe-lamp, used as an excitation source, a polarization state generator (PSG) and a polarization state analyzer (PSA) unit to generate and analyze the required polarization states, coupled to an imaging spectrograph/ CCD assembly for spectrally resolved signal detection. The broadband light emitted by a Xe-lamp source (HPX-2000, Ocean Optics, USA) was collimated using a combination of lenses, passed through the PSG unit and illuminated the tissue sample with a spot size ~ 1-mm-diameter. The PSG unit consist of a linear polarizer (P1, LPVIS100, Thorlabs, USA) with its axis oriented along the laboratory horizontal direction, followed by a rotatable achromatic quarter waveplate ($QWP_1$, WPQ10M-633, Thorlabs, USA) mounted on a computer-controlled rotational mount (PRM1/MZ8, Thorlabs, USA). The backscattered light from the sample was collected by a lens, passed through the PSA unit and relayed to an imaging spectrograph (Shamrock, SR- 303i-A, USA) coupled to a CCD detector (ANDOR technology, UK). The PSA unit essentially comprises of a similar arrangement of fixed linear polarizer (P2, oriented at vertical position) and a rotatable achromatic quarter wave retarder ($QWP_2$), but positioned in a reverse order. The entrance slit of the imaging spectrograph was kept at a focal distance of the collection lens. For this arrangement, the angular distribution of the backscattered light (with collection half angle ~ 3º around the exact backscattering angle $\theta = 180º$), for a fixed azimuthal angle, was projected to the spectrograph entrance slit [13]. The spectra (wavelength 400 – 800 nm) were acquired with a spectral resolution of ~ 1 nm.

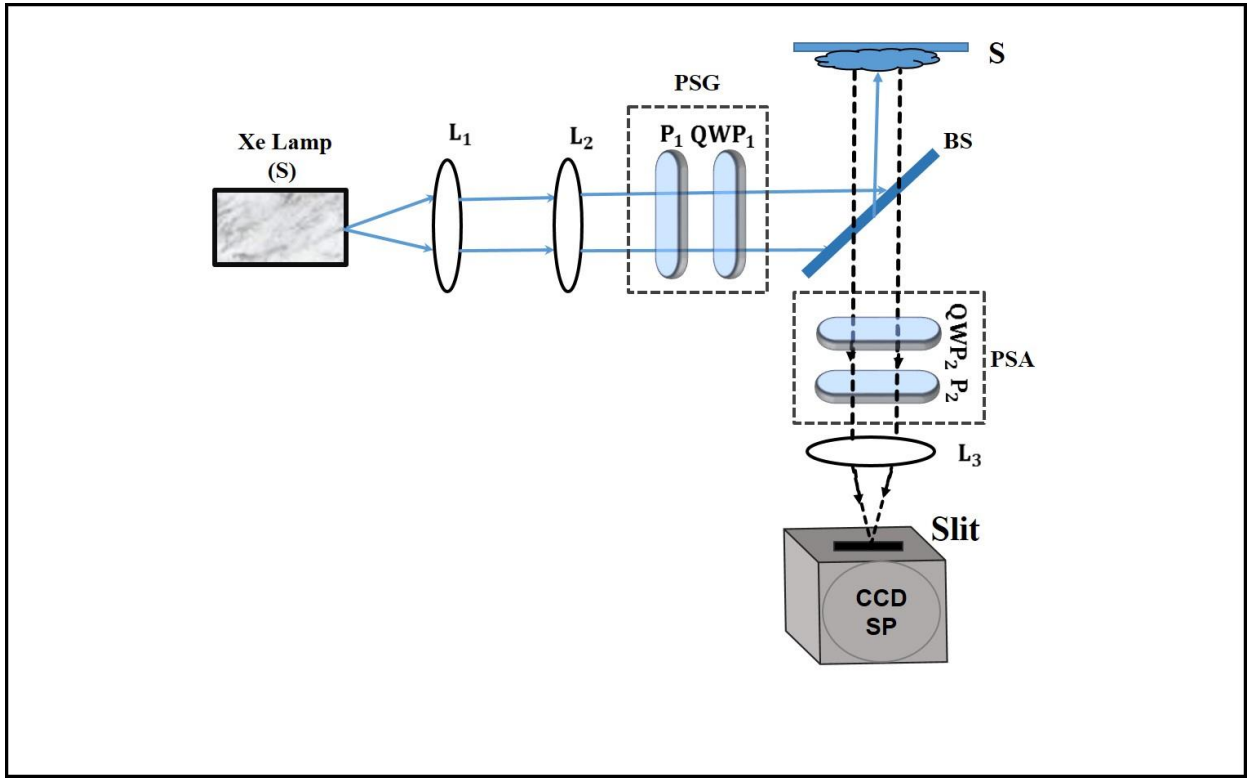

**Figure 1:** Schematic of the experimental system for backscattering spectroscopic Mueller matrix measurement. Xe Lamp: Excitation Source; $L_1$, $L_2$, $L_3$: Collimating Lenses; $A_1$: Aperture; PSG: Polarization State Generator ($P_1$: Linear Polarizer, $QWP_1$: Quarter Wave Plate); S: Tissue sample; BS: Beam Splitter; PSA: Polarization State Analyzer ($P_2$: Linear Polarizer, $QWP_2$: Quarter Wave Plate); $L_4$: Collecting Lens; CCD SP: CCD Spectrograph.

The specifics of the Mueller matrix measurement strategy can be found elsewhere [29]. Briefly, the 4×4 spectral Mueller matrices are constructed by combining sixteen sequential spectrally resolved intensity measurements (spectra) for four different combinations of the optimized elliptical polarization state generator (using the PSG unit) and analyzer (using the PSA unit) basis states. The four elliptical polarization states are generated by sequentially changing the fast axis of $QWP_1$ to four angles ($\vartheta$ =35º, 70º, 105º and 140º) with respect to the axis of P1. Similarly, the four elliptical analyzer basis states are obtained by changing the fast axis of $QWP_2$ to the corresponding four angles (35º, 70º, 105º and 140º). These sixteen polarization-resolved scattering spectra are combined to yield the scattering Mueller matrix of the sample (see Supplementary information).

A differential interference contrast (DIC) microscope was used to measure the spatial variation of tissue RI. Images were recorded at 60X magnification and with an width of point spread function ~ 0.36 µm. Histopathologically characterized (precancerous Grade I, II and III) biopsy samples of human cervical tissues (lateral dimension 4 mm × 6 mm, thickness ~ 50 µm) were obtained from G.S.V.M. Medical College and Hospital, Kanpur, India.

## 4. Results and Discussion

### *4.1 Signature of multifractal anisotropy in spatial variation of tissue RI*

We unfolded (pixel-wise) the DIC image (containing 256×256 square pixels) along two orthogonal (horizontal and vertical) directions (yielding a spatial series of length $N = 2^{16}$) and then subjected the resulting two 1D RI fluctuation series to MFDFA. The results of MFDFA analysis on the connective tissue regions of a typical Grade-I precancerous cervical tissue is summarized in **Figure 2**. The observed wide range of fluctuations (length scale ranging from sub-micron to tens of microns) underscores the overall randomness of the index variations

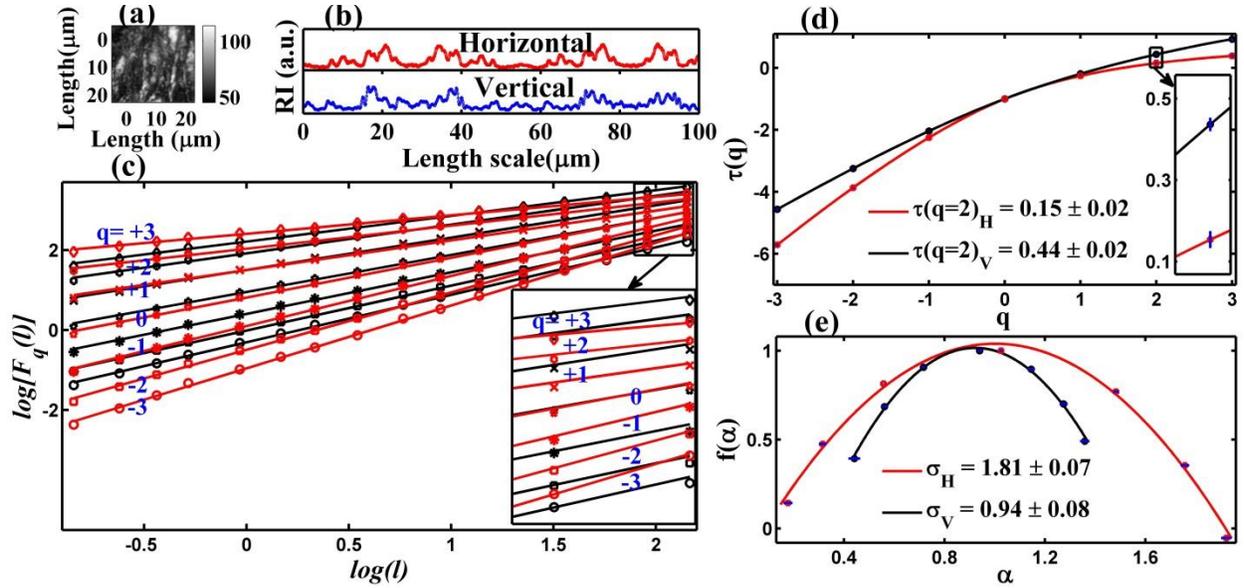

**Figure 2.** Signature of multifractal anisotropy in spatial variation of tissue RI. a) DIC image of typical Grade I precancerous connective tissue. (b) The detrended RI fluctuations for horizontal (top panel) and vertical (bottom panel) unfolding of typical Grade I precancerous connective tissue. The X-axis represents the length scale (in μm). (c) The log-log (natural logarithm) plot of the moment (q = -3 to +3) dependent fluctuation function $F_q(l)$ vs length scale $l$ for horizontally (red line and symbols) and vertically (black lines and symbols) unfolded RI fluctuations. The inset highlights the differences in the slopes between them. (d) The moment ($q$)-dependence of the classical scaling exponent $\tau(q)$ (inset highlights the anisotropy or difference in $\tau$ around $q = 2$) and (e) the singularity spectra $f(\alpha)$ for horizontally (red square) and vertically (black circle) unfolded RI fluctuations. The values for $\tau(q = 2)$ and width of the singularity spectra ($\sigma$) are noted. In (d) and (e) lines are guide for eye and the error bars represent standard deviations of the parameters for measurements on twenty non-overlapping sites.

[Figure 2(b), for horizontal and vertical unfolding]. Statistical self-similarity is manifested in the spatial frequency (ν) distribution of the Fourier power spectrum (see Supplementary **Figure S1**), which exhibited power-law scaling beyond a certain ν range (ν ≥ 0.033 μm$^{-1}$). The length scale ($L_o \sim \nu_{min}^{-1} \sim$ 30 μm) is regarded as the fractal upper scale [19]. The power law coefficient (slope $\gamma$) was however, not uniform throughout the entire ν range (Figure S1), underlining the complex nature of the spatial correlation and the resulting scaling behavior. The MFDFA-derived (using

Equation 7) $F_q(l)$ vs length scale (*l*) plots (shown in log-log scale in Figure 2(c)) furnish evidence of multifractality, as the slopes ( $F_q(l) \sim l^{h(q)}$) vary significantly with varying moment (*q*). This analysis was performed within length scales limited by the width of the PSF of the microscope (~ 0.4 µm) and the fractal upper scale. Moreover, the analysis was also restricted for values $q$ between $-3\ to +3$ [30]. Strength of multifractality is subsequently quantified (using Equations 8 and 9) via the classical scaling exponent $\tau(q)$ [Figure 2(d)] and the width ($\sigma$) of the singularity spectrum $f(\alpha)$ [Figure 2(e)]. For the sake of statistical independence, the measurements and analysis was performed on ten non-overlapping sites (DIC images) of the same tissue sample. The $\tau(q)$ and the $f(\alpha)$ spectra shown in the figure represent the mean values and the corresponding standard deviations are displayed. Strong multifractality is evident from significant deviation of $\tau(q)$ from the linear behavior and corresponding considerable magnitudes of $\sigma$, for either horizontal or vertical unfolding. Interestingly, the *q*-dependent slopes of the $log[F_q(l)]$ vs $log(l)$ plots are different for horizontal and vertical unfoldings (Figure 2(b)), which is indications of the anisotropic nature of multifractality. This multifractal anisotropy was also statistically significant as the trend was consistent for all the ten non-overlapping tissue sites (see Supplementary Figure S2 where $log[F_q(l)]$ vs $log(l)$ plots for *q*=2 are shown including the standard deviations). The corresponding signature of multifractal anisotropy is reflected as prominent differences in the values of $\tau(q = 2)$ and $\sigma$ between the horizontal ($\tau(q = 2) = 0.15 \pm 0.02$, $\sigma = 1.81 \pm 0.07$) and the vertical ($\tau(q = 2) = 0.44 \pm 0.02$, $\sigma = 0.94 \pm 0.08$) index fluctuations (Figure 2(d) and 2(e))).

Connective tissue is comprised of collagen fiber network, formed via complex hierarchical organization of the building blocks – collagen molecules, micro-fibrils and macroscopic fiber bundles [31]. Anisotropic organization of these building blocks may manifest as multifractal anisotropy [32]. Interestingly, the differences in $\tau(q)$ values between the horizontal and vertical index fluctuations are more prominent for negative $q$-values [Figure 2(d)]. Since, negative (positive) values of $q$ primarily capture signatures of small (large) scale fluctuations, this highlights the importance of the small scale index variations in the resulting anisotropy (possibly originating from anisotropic organization at the sub-micron length scale [32]).

*4.2 Quantification of multifractal anisotropy from scattering Mueller matrix: Inverse analysis*

Making use of the experimental system (see Figure1), backscattering spectroscopic Mueller matrices were recorded (**Figure 3**(a)) from a Grade-I precancerous cervical tissue (corresponding to Figure 2). Several interesting trends can be gleaned from the Mueller matrix (see Figure 3(a)). The considerably low magnitudes of $M_{24}(\lambda)/M_{42}(\lambda)$, $M_{34}(\lambda)/M_{43}(\lambda)$, $M_{32}(\lambda)/M_{23}(\lambda)$ elements indicate weak *macroscopic* linear retardance effect, which is likely due to the random macroscopic organization of the collagen fibrous network [17,31]. In contrast, the $M_{12}(\lambda)$ and $M_{13}(\lambda)$ elements and the resulting linear diattenuation parameter $d(\lambda)$ (Figure 3(b)) exhibit considerable magnitude and wavelength variation (decays with λ). As envisaged, this is an exclusive signature of fractal (multifractal) anisotropy, which is subsequently manifested in the spatial frequency ($\nu = \frac{2}{\lambda}\sin\left(\frac{\theta}{2}\right), \theta = 180°$) distribution of the derived (by using Equation(3)) light scattering parameters, $|s_1(\nu)|^2$ and $|s_2(\nu)|^2$ (Figure 3(c)). When fitted with Equation(4), they yield multiple power law exponents at different ν-ranges. Importantly, fitting over broad ν-range yielded differences in average power law exponents; $\gamma_{||}$= -4.1 and $\gamma_{\perp}$= -4.3 derived from $|s_2(\nu)|^2$ and $|s_1(\nu)|^2$, respectively ($\Delta\gamma = |\gamma_{||} - \gamma_{\perp}| = 0.2$). These results provide further evidence of anisotropic self-similarity in spatial RI fluctuations.

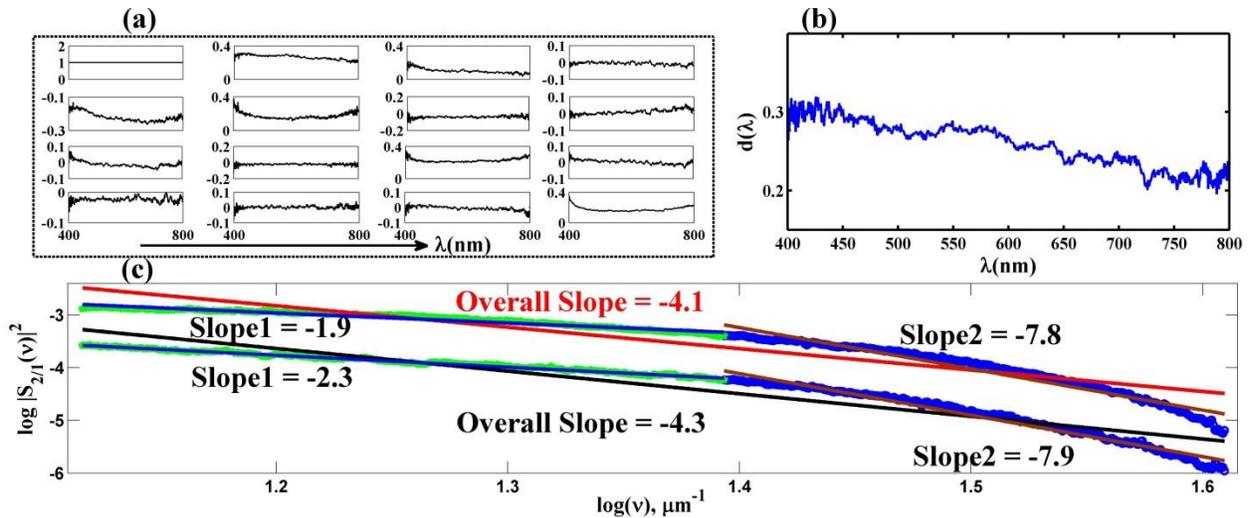

**Figure 3.** Manifestation of multifractal anisotropy in the wavelength variation of scattering Mueller matrix elements of a Grade I precancerous tissue [corresponding to Figure 2]. (a) Wavelength variation of the scattering Mueller matrix elements (norma,ized by the $M_{11}(\lambda)$ element). (b) Wavelength dependence of the derived linear diattenuation parameter $d(\lambda)$. (c) The spatial frequency ($\nu$) distribution of the Mueller matrix-derived light scattering parameters $|s_2(\nu)|^2$ and $|s_1(\nu)|^2$ (log-log plot). Fitting at two selected $\nu$-ranges (lower (blue) and higher (brown)) and overall fitting (red for $|s_2(\nu)|^2$ and black for $|s_1(\nu)|^2$) are shown and the corresponding values for the exponents $\gamma$ are noted.

**Figure 4** displays the results of the *inverse* analysis performed on $|s_2(\nu)|^2$ and $|s_1(\nu)|^2$. The derived (by using Equation (5)) index inhomogeneity distributions $\eta_{||}(\rho)$ and $\eta_{\perp}(\rho)$ encode information on index inhomogeneities with length scales down to sub-micron level (see Figure 4(a)). Multifractality is once again evident from the large variations in the slopes of *log $F_q(l)$ vs log l* with varying *q*, derived from either $\eta_{||}(\rho)$ and $\eta_{\perp}(\rho)$ (see Figure 4(b)). The strengths of multifractality are subsequently quantified via $\tau(q)$ (Figure 4(c)) and width of $f(\alpha)$ (Figure 4(d)). Here the $\tau(q)$ and the $f(\alpha)$ spectra represent the mean values over ten neighboring sites of the same tissue sample, and the corresponding standard deviations are also shown. Significant magnitudes of the differential classical scaling exponent $[\Delta\tau = |\tau(q=2)_{||} - \tau(q=2)_{\perp}| = 0.16]$ and differential width of singularity spectrum $[\Delta\sigma = |\sigma_{||} - \sigma_{\perp}| = 0.85]$ provide

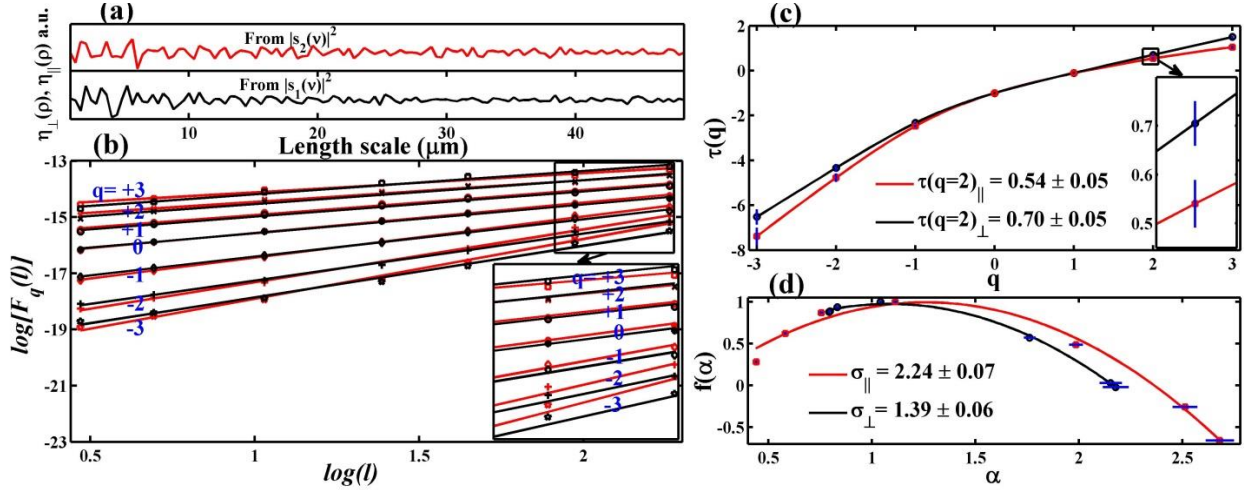

**Figure 4.** The results of the *inverse* analysis performed on $|s_2(v)|^2$ and $|s_1(v)|^2$ light scattering parameters [corresponding to Figure 3]. (a) The derived index inhomogeneity distributions $\eta_{||}(\rho)$ (from $|s_2(v)|^2$, top panel) and $\eta_{\perp}(\rho)$ (from $|s_1(v)|^2$, bottom panel). The X-axis represents the statistically equivalent length scale (in μm). (b), (c), (d)**:** The results of MFDFA inverse analysis on $\eta_{||}(\rho)$ and $\eta_{\perp}(\rho)$. (b) Log-log plot of the moment (q = -3 to +3) dependent fluctuation function $F_q(l)$ vs $l$ derived from $\eta_{||}(\rho)$ (red line and symbols) and $\eta_{\perp}(\rho)$ (black line and symbols) (c) The moment ($q$)-dependence of the classical multifractal scaling exponent $\tau(q)$ (inset highlights the anisotropy or difference in $\tau$ around $q = 2$) and (d) the corresponding singularity spectra $f(\alpha)$ derived from $\eta_{||}(\rho)$ (red square) and $\eta_{\perp}(\rho)$(black circle). The values for $\tau(q = 2)$ and $\sigma$ are noted. In (c) and (d) lines are guide for eye and the error bars represent standard deviations of the parameters for measurements on ten non-overlapping spots.

conclusive evidence of multifractal anisotropy. Although a direct quantitative comparison of the parameters determined from the Mueller matrix *inverse* analysis (see Figure4) and from *forward* analysis on the DIC image (Figure 2) is not feasible, the observed trends show self-consistency. Specifically, in both cases (Figure 2(d) and Figure 4(c)), multifractal anisotropy is manifested as a difference in $\tau(q)$ values between the two orthogonal index inhomogeneity distributions for negative *q*-values. As previously noted, these imply that the multifractal anisotropy parameters capture morphological information on tissue micro-structural anisotropy (sub-micron length scales).

**Figure 5** summarizes the results of exploration the *multifractal anisotropy parameters* for differentiating different grades of precancers. Results of twenty three tissues (Grade I -11, Grade II- 6 and Grade III – 6) demonstrate that when mapped by $\Delta\tau$ and $\Delta\sigma$ parameters, the three different grades can be differentiated. Here, the site-averaged multifractal parameters from each of the tissue samples are shown. The values of $\Delta\tau$ and $\Delta\sigma$ decreases with increasing pathology grades, suggesting a reduction in the multifractal anisotropy. The observed drastic reduction in $\Delta\sigma$ (as compared to $\Delta\tau$) underscores the fact that the changes in multifractal anisotropy with increasing pathology grades are primarily related to subtle morphological alterations at the microscopic domain of connective tissue [31,32]. This follows because the singularity spectrum $f(\alpha)$ utilizes the entire $\tau(q)$ spectrum (as opposed to $\tau(q=2)$) including the negative $q$ region (which capture small length scale fluctuations) where the local anisotropy effect is manifested more prominently (Figure 4(c)). The reduction in multifractal anisotropy may thus originate from disorganization of locally anisotropic microscopic domains (the collagen molecules and micro-fibrils) and / or reduction in local microscopic birefringence [31,32]. This hypothesis was verified on extracted Bovine collagen samples treated with 5% acetic acid (see Supplementary information). In contrast, none of the above trends could be gleaned from Mueller matrix-derived conventional *macroscopic* linear retardance δ parameters, which were generally low for all the samples and did not exhibit appreciable difference between different grades (e.g., δ at 600 nm were *0.08± 0.03, 0.04 ± 0.02 and 0.09±0.05 rad* for Grade I, Grade II and Grade III tissues). As previously discussed, complex organization of the fibrous network and the resulting *macroscopically* random orientation of the anisotropic domains eventually manifested as weak macroscopic linear retardance effect.

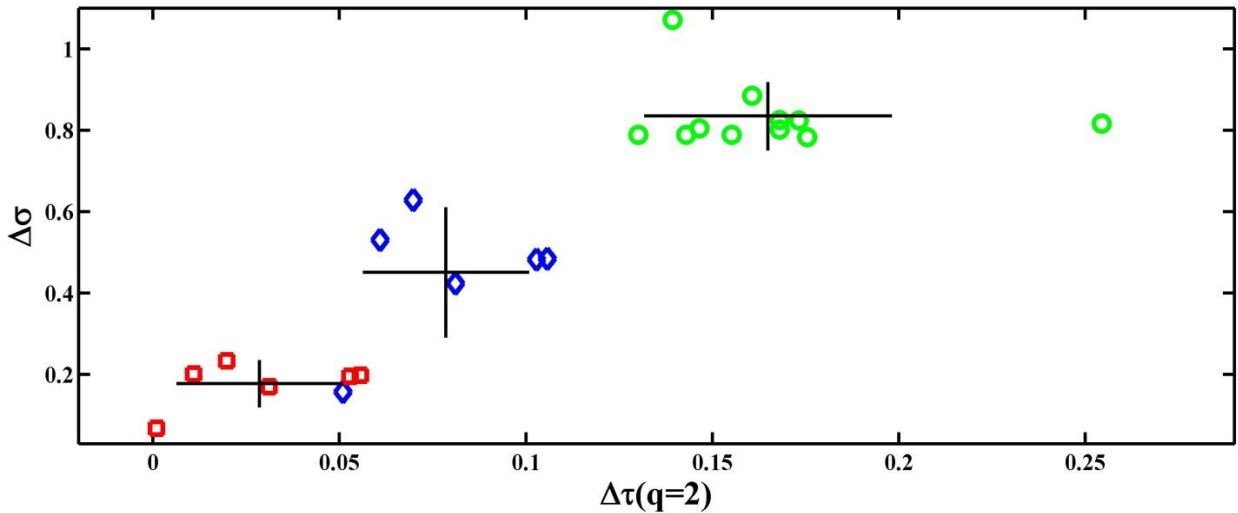

**Figure 5.** Differentiating different grades of precancers based on the Mueller matrix-derived *multifractal anisotropy parameters*. The three different precancerous grades (Grade I – green circle, II – blue diamond and III – red square) are mapped by their differential classical multifractal scaling exponent, $\Delta\tau = |\tau(q=2)_{\|} - \tau(q=2)_{\perp}|$; and differential width of singularity spectrum $\Delta\sigma = |\sigma_{\|} - \sigma_{\perp}|$ for orthogonal linear polarizations. Higher grades of precancers are associated with decrease of both $\Delta h$ and $\Delta\sigma$ parameters, implying reduction in multifractal anisotropy.

## 5. Conclusions

To summarize, we have demonstrated that the spatial variation of refractive index in biological tissue exhibit multifractal anisotropy, leaving its exclusive signature as an intriguing spectral linear diattenuation effect in the scattering Mueller matrix. We have accordingly introduced a new set of *multifractal anisotropy* parameters, the differential classical scaling exponent ($\Delta\tau$) and the differential width of the singularity spectrum ($\Delta\sigma$) and developed an *inverse* analysis method for their quantification from spectral Mueller matrix. These parameters appear sensitive to structural anisotropy at the micron/sub-micron length scales. The method was initially explored for detecting cervical precancerous alterations on *ex vivo* tissues, with early indications showing promise. The ability to probe and quantify subtle changes in tissue micro-structural anisotropy using polarimetric backscattering measurements bodes well for *in vivo* deployment because (i) the required Mueller matrix elements can be obtained using linear polarization measurements alone, and (ii) the backscattering geometry is clinically amenable. Finally, the novel ability to sense structural anisotropy in the sub-micron length scale via the newly defined multifractal anisotropy parameters may open the door for non-invasive characterization of a variety of complex materials and disordered systems.

**Acknowledgements**

This work was supported by IISER-Kolkata, an autonomous institute funded by MHRD, Govt. of India. We would also like to acknowledge BRNS-DAE, Govt. of India for funding. The authors thank Dr. Asha Agarwal, G.S.V.M. Medical College and Hospital, Kanpur, for providing the histopathologically characterized tissue samples.

*Supplementary Information*

**A. Mueller matrix measurement scheme**

The Mueller matrix measurement strategy is based on sixteen spectrally resolved intensity measurements performed by sequentially generating and analyzing four elliptical polarization states using a polarization state generator (PSG, comprising of a fixed linear polarizer $P_1$ with its axis oriented along the laboratory horizontal direction, followed by a rotatable quarter wave retarder $QWP_1$) and a polarization state analyzer (PSA, similar arrangement of fixed linear polarizer $P_2$ with its axis oriented along vertical direction and rotatable quarter wave retarder $QWP_2$, but positioned in a reverse order) unit respectively. The four required (and optimized) elliptical polarization states are generated by sequentially orienting the axis of the quarter wave retarder $QWP_1$ to four optimized angles, 35º, 70º, 105º and 140º with respect to the polarizer axis [1]. These four generated polarization states can be represented by a 4 × 4 matrix *W*, whose column vectors are the corresponding four generated Stokes vectors. After sample interaction (backscattered from tissue sample), the resulting output polarization states of the backscattered light are analyzed in the PSA unit, by sequentially changing the orientation angle of $QWP_2$ to the same angles as PSG (35º, 70º, 105º and 140º). The PSA results can also be described by a $4 \times 4$ analyzer matrix $A$. The Stokes vectors of the light to be analyzed are projected onto the four basis states, given by the rows of $A$. The sixteen intensity measurements required for the construction of a full Mueller matrix are grouped into the measurement matrix $M_i$, which can be related to PSA/PSG matrices $W$ and $A$, as well as the sample Mueller matrix $M$ by [1]:

$$M_i = AMW. \qquad (S1)$$

Once the exact forms of the A and W matrices are known, the sample Mueller matrix can be determined as

$$M = A^{-1}MW^{-1} \qquad (S2)$$

calibrating samples having known forms of Mueller matrices (e.g. a set of quarter wave plates and polarizers) [1].

The exact experimental forms of the $W$ and $A$ matrices and their wavelength dependence were determined using the so-called Eigenvalue calibration method by performing measurements on this approach enabled us to correct for non-ideal behaviour of optical components, misalignments etc., and ensured high accuracy (accuracy ~ 0.01 in normalized matrix elements) of Mueller matrix measurements over the entire spectral range 400 – 800 nm. The details of the eigenvalue calibration method can be found in our previous publication [1] and elsewhere [2].

**B. Determination of the macroscopic polarization anisotropy parameters of tissue**

In order to estimate the medium polarimetry parameters, the experimental Mueller matrix $M$ was decomposed into basis matrices of the three possible polarimetry effects [3]:

$$M \Leftarrow M_\Delta \cdot M_R \cdot M_D. \qquad (S3)$$

With $\Leftarrow$ symbol used to signify the decomposition process. Here, the matrix $M_\Delta$ describes the depolarizing effects of the medium, $M_R$ accounts for the effects of linear and circular retardance (or optical rotation), and $M_D$ includes the effects of linear and circular diattenuation. The two parameters pertinent to this study, the linear diattenuation $d$ and linear retardance $\delta$ were derived from the decomposed matrices using standard procedure as [3]:

$$d = \frac{1}{M_{11}} \sqrt{M_{12}^2 + M_{13}^2}$$

$$\delta = \cos^{-1}\left\{ \sqrt{[M_R(2,2) + M_R(3,3)]^2 + [M_R(3,2) - M_R(2,3)]^2} - 1 \right\}. \qquad (S4)$$

**C. Results of Fourier and MFDFA analysis on DIC images of tissue**

**C.1. Results of the Fourier power spectrum analysis on DIC images of tissue**

The results of MFDFA analysis on the DIC image of connective tissue regions of a typical Grade-I precancerous human cervical tissue was shown in Figure 2 of the manuscript. The horizontally and the vertically unfolded RI fluctuations were displayed in Figure 2(a). Here, we show the corresponding Fourier power spectrum in Supplementary Figure S1. Note that for one dimensional fluctuation series exhibiting statistical self-similarity, (such as the 1D RI fluctuation series obtained by unfolding the DIC image) the Fourier power spectrum assumes a form of power law $(P(\nu) \sim \nu^{-\gamma}$, $\nu$ being the spatial frequency here in $\mu m^{-1}$) at the limit of large frequencies $\nu$. As evident, statistical self-similarity is manifested as a power-law scaling of the spatial frequency distribution in the Fourier power spectrum (Figure S1(a) and S1(b) for horizontal and vertical unfolding, respectively), which exhibit power law scaling beyond a certain spatial frequency $\nu$ range (for $\nu \geq 0.033$ $\mu m^{-1}$, the spectral density appeared linear on a

log–log plot). The length scale ($L_o \sim \nu_{min}^{-1} \sim 30$ μm) can accordingly be regarded as the fractal upper scale. The power law exponent (slope $\gamma$) is however, not uniform throughout the entire $\nu$ range. Fitting at two selected $\nu$-ranges (lower (green) and higher (red), respectively) and overall fitting (black) yield different values for slopes. Use of the overall slope (fitting over broad $\nu$-range) yielded differences in the average power law exponents, $\gamma$ = *1.44* and *1.70* for horizontally and vertically unfolded index fluctuations, respectively ($\Delta\gamma$ = 0.26). These results provide initial indications of anisotropic nature of self-similarity in spatial RI fluctuations.

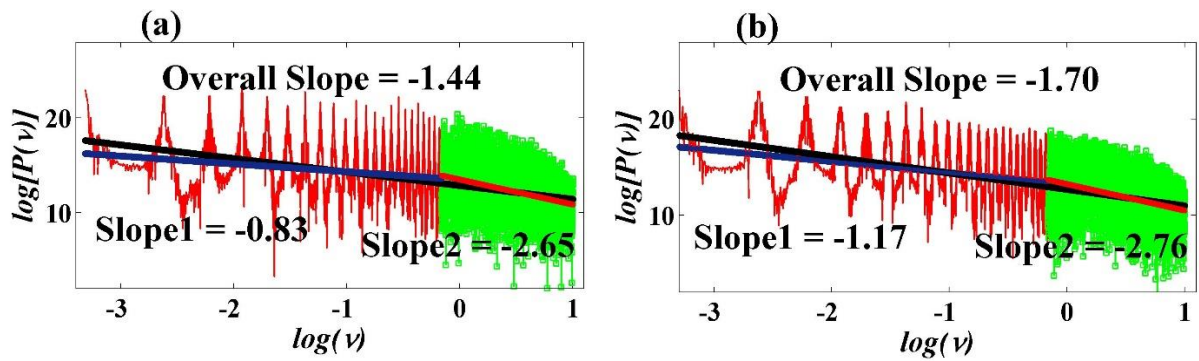

**Supplementary Figure S1:** Fourier power spectrum of the generated 1D RI fluctuation series (corresponding to Figure 1 of the manuscript) for (a) horizontal unfolding and (b) for vertical unfolding (shown in natural logarithm scale). Fitting at two selected $\nu$-ranges (lower (blue) and higher (red), respectively) and overall fitting (black) yield different values for slopes (power law exponents $\gamma$) and the corresponding values for $\gamma$ are noted.

**C.2 Manifestation of multifractal anisotropy in the MFDFA-derived fluctuation functions**

The results of the MFDFA analysis on the DIC images (horizontally and vertically unfolded to yield two 1D RI fluctuation series) connective tissue regions of a typical Grade-I precancerous connective tissue was shown in Figure 2 of the manuscript. The corresponding fluctuation functions $F_q(l)$ (derived using Equation 7) for both horizontal and vertical unfolding were displayed in Figure 2(b). Here, we separately display the log-log (natural logarithm) plot of the fluctuation functions $F_q(l)$ vs length scale $l$ for moment (q = 2) only (Supplementary Figure S2(a)). The results are shown for ten non-overlapping sites (DIC images) of the same tissue sample. The mean values of $F_q(l)$ and their standard deviations (shown by error bars) are displayed. Power law behaviour of $F_q(l)$ is evident (appears linear in log-log plot) and the corresponding slopes or the generalized Hurst exponents $h(q = 2)$ (assuming $F_q(l) \sim l^{h(q)}$) are clearly different for horizontal and vertical unfolding ($h(q = 2) = 0.58 \pm 0.01$ for horizontal unfolding as compared to $h(q = 2) = 0.72 \pm 0.01$ for vertical unfolding. These results provide conclusive evidence that the observed multifractal anisotropy trends in spatial RI fluctuations of human connective tissues are consistent and statistically significant. We also additionally show results of similar analysis on a typical Grade-III precancerous connective tissue in Supplementary Figure S2(b). The reduction in multifractal anisotropy is evident.

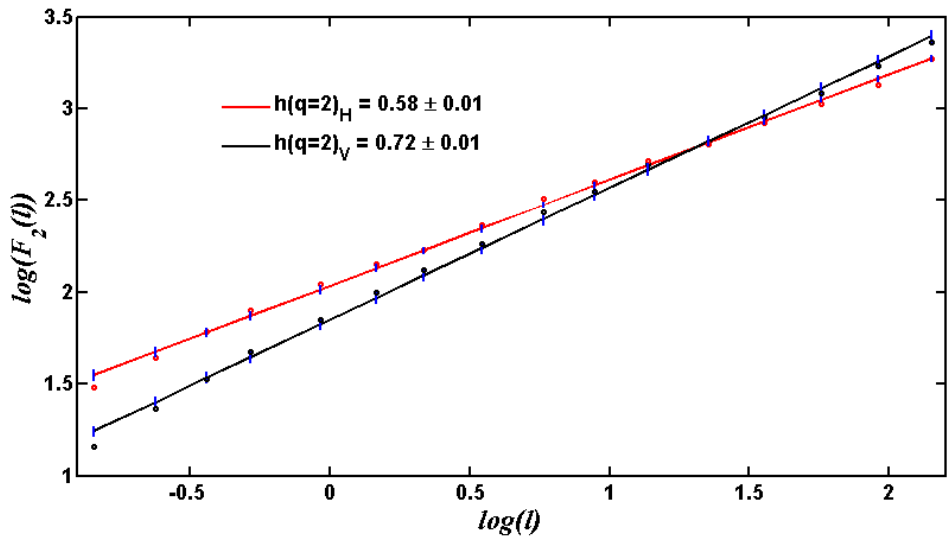

**Supplementary Figure S2a:** Log-log plot of fluctuation function $F_{q=2}(l)$ vs length scale $l$ derived (using Equation 7) for horizontally (red line and symbols) and vertically (black lines and symbols) unfolded RI fluctuations of a Grade I precancerous cervical connective tissue (corresponding to Figure 1). The values for the slopes, the generalized Hurst exponents $h(q = 2)$ are noted. The error bars represent standard deviations for measurements on ten non-overlapping sites.

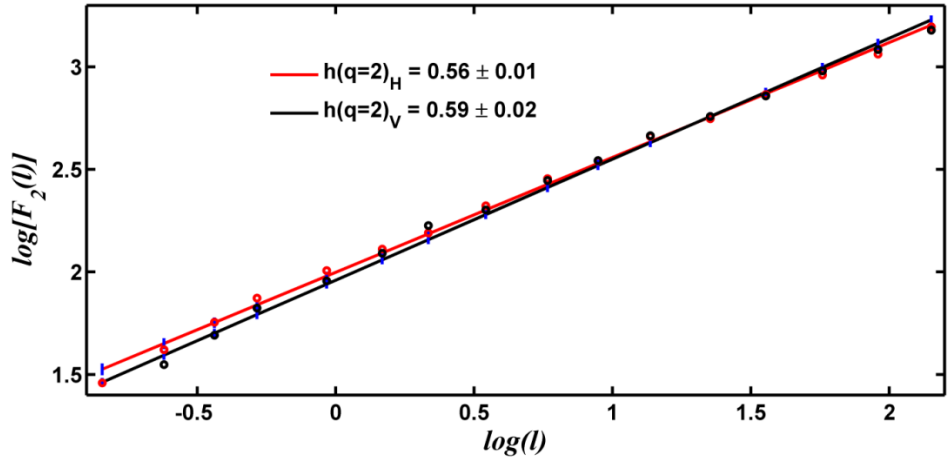

**Supplementary Figure S2b:** Log-log plot of fluctuation function $F_{q=2}(l)$ vs length scale $l$ derived (using Equation 7) for horizontally (red line and symbols) and vertically (black lines and symbols) unfolded RI fluctuations of a Grade III precancerous cervical connective tissue. The values for the slopes, the generalized Hurst exponents $h(q=2)$ are noted. The error bars represent standard deviations for measurements on ten non-overlapping sites. Reduction in multifractal anisotropy (as compared to the Grade-I tissue of Figure S3(a)) is evident.

**C.3 Validation of the multifractal analysis on monofractal fluctuation series generated using fractional Brownian motion algorithm**

In order to test the accuracy of our MFDFA analysis to extract the multifractal parameters of spatial RI fluctuations of tissue (unfolded DIC image), we first computed monofractal fluctuation series using fractional Brownian motion (fBm) algorithm [4], and then subjected the fluctuation series to MFDFA. The fBm series was generated for the same statistics (data length and interval) as our experimental spatial RI fluctuation series. For this purpose, the input Hurst exponent

($H$ (0,1)) was taken corresponding to the Fourier power spectrum of the 1D RI fluctuation series in monofractal approximation (corresponding to Figure S1). Note that in monofractal approximation, the power law coefficient of the power spectrum $\gamma$ ($P(\nu) \sim \nu^{-\gamma}$), is related to the Hurst exponent $H$ as $\gamma = 2H + 1$ [5]. Accordingly, the input values for $H$ were chosen to be $H = 0.22$ (corresponding to horizontal unfolding Figure S2a) and $H = 0.35$ (corresponding to vertical unfolding Figure S2(b)). The two generated fBm series were subsequently subjected to both Fourier analysis and MFDFA analysis and the results are presented in Supplementary Figure S3. Unlike the power spectra of our experimental RI fluctuations (Figure S1), the power spectra of the fBm-generated fluctuation series do not exhibit multiple power law exponents (fitting at different selected $\nu$-ranges, did not yield significantly different values of the slopes) and the derived power law exponents were also close to the expected input values. Importantly, the MFDFA-derived generalized Hurst exponents $h(q)$ did not exhibit any appreciable dependence on the moment ($q$) and these were observed to obey the expected behaviour $h(q = 2) \sim H$. These results confirm that the MFDFA-derived multifractal trends of our experimental tissue RI fluctuations originated from 'true' multifractality in the spatial variations of tissue RI.

The relationship between the scaling exponents used in standard multifractal analysis [6] and those in the MFDFA approach adopted by us, worth a brief mention here. As shown by Kantelhardt et al [7], the classical multifractal scaling exponent $\tau(q)$ defined in MFDFA is same as that defined in partition function-based standard multifractal formalism [7,8]. Moreover, $\tau(q)$ is related to the scaling exponent $\xi(q)$ of structure function-based approaches as $\xi(q) = 1 + \tau(q)$ [6,8]. Finally, as described in the manuscript, for monofractal series $\tau(q) = qH - 1$, where $H$ is the global Hurst exponent [$H \in (0,1)$]. Moreover, the scaling exponent $\tau(q=2)$ is related to the power law exponent of the Fourier power spectrum as $\gamma = \tau(q=2) + 2$ [9].

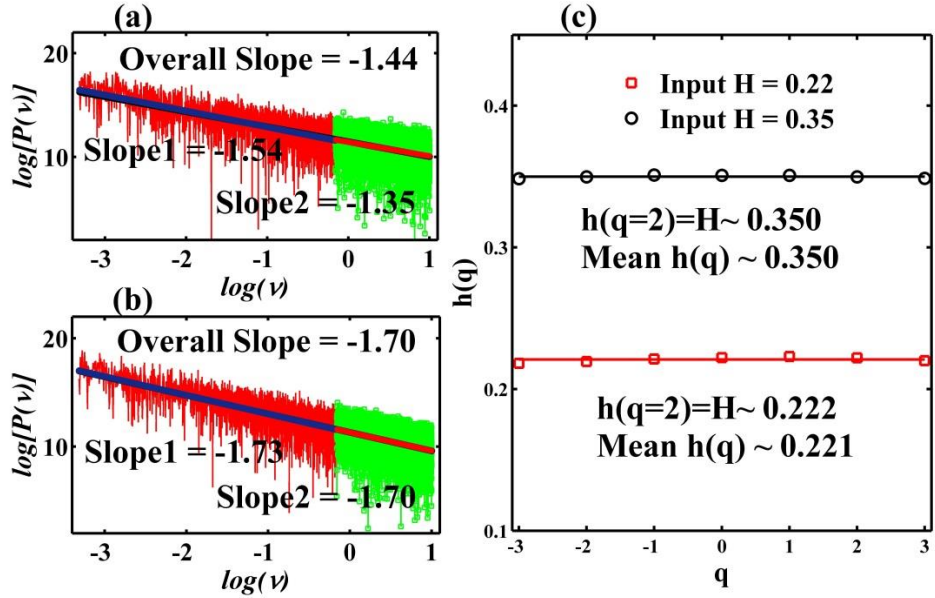

**Supplementary Figure S3:** Fourier power spectra (shown in natural logarithm scale) of the fluctuation series generated using Fractional Brownian motion (fBm) algorithm. The fluctuation series were generated using input values of Hurst exponents of H = 0.22 (corresponding power spectrum shown in (a)) and 0.35 (power spectrum shown in (b)) and by using the same statistics (data length and interval) as our experimental data (corresponding to Figure S2 and Figure 1 of the manuscript). Fitting at two selected ν-ranges (lower (blue) and higher (red), respectively) and overall fitting (black) do not yield different values for the slopes. (c) The MFDFA-derived moment ($q$)-dependence of the generalized Hurst exponent $h(q)$ for the same fluctuation series.

We would like to note that the average power law exponents $\gamma$ obtained from the Fourier power spectrum of our experimental 1D RI fluctuation series (supplementary Figure S1) are observed to be slightly different from that expected using the MFDFA-derived classical scaling exponents

$\tau(q = 2)$ (Figure 1(c) of the revised manuscript). Note that the complex nature of the spatial RI fluctuations yielded a power spectrum whose power law exponent $\gamma$ was not uniform throughout the entire ν range (fitting at different selected ν-ranges yielded different values for slopes, Figure S1). Even though, we estimated the average value for $\gamma$ by fitting the power spectrum over a broad range of spatial frequency ν, this was associated with relatively large standard deviation of the fitted parameter. In such situation, it may be difficult to extract and assign a unique power law exponent which can be directly linked to $\tau(q = 2)$ via $= \tau(q = 2) + 2$. It is also pertinent to note that the power law coefficients of the light scattering parameters $|s_2(\nu)|^2$ and $|s_1(\nu)|^2$ (Figure 2 of the manuscript) are different by a factor ~ 2 from that of the power spectrum of the 1D RI fluctuation series (obtained from the DIC image). This is due to the fact that the latter being obtained from 1D Fourier transform of the index fluctuation, while the former emerges via a 3D Fourier transform of the scattering potential (in Born approximation of light scattering).

### D. Results on Bovine collagen samples

The spectral scattering Mueller matrices were recorded from Bovine collagen samples with and without acetic acid (5 %) treatment. The inverse analysis was performed on the spectral variation of the scattering Mueller matrix elements using Equation (5) of the manuscript, to yield the index inhomogeneity distributions $\eta_{||}(\rho)$ and $\eta_{\perp}(\rho)$. These were subsequently analyzed using MFDFA. The results are summarized in Supplementary Figure S4.

Acetic acid treatment is known to break the collagen molecular cross-links, leading to disorganization at the microscopic level [10]. Reduction in the magnitudes of the multifractal anisotropy parameters (differential classical scaling exponent, $\Delta\tau = |\tau(q = 2)_{||} - \tau(q = 2)_{\perp}|$ and differential width of singularity spectrum $\Delta\sigma = |\sigma_{\parallel} - \sigma_{\perp}|$ for orthogonal linear polarizations) following acetic acid treatment, provide initial evidence that the multifractal anisotropy parameters are able to probe structural anisotropies at the microscopic domain. These also support our hypothesis on the observed trends on precancerous tissue samples: the reduction in multifractal anisotropy may originate from the disorganization of locally anisotropic microscopic domains (the collagen molecules and the micro-fibrils) and / or reduction in local microscopic birefringence.

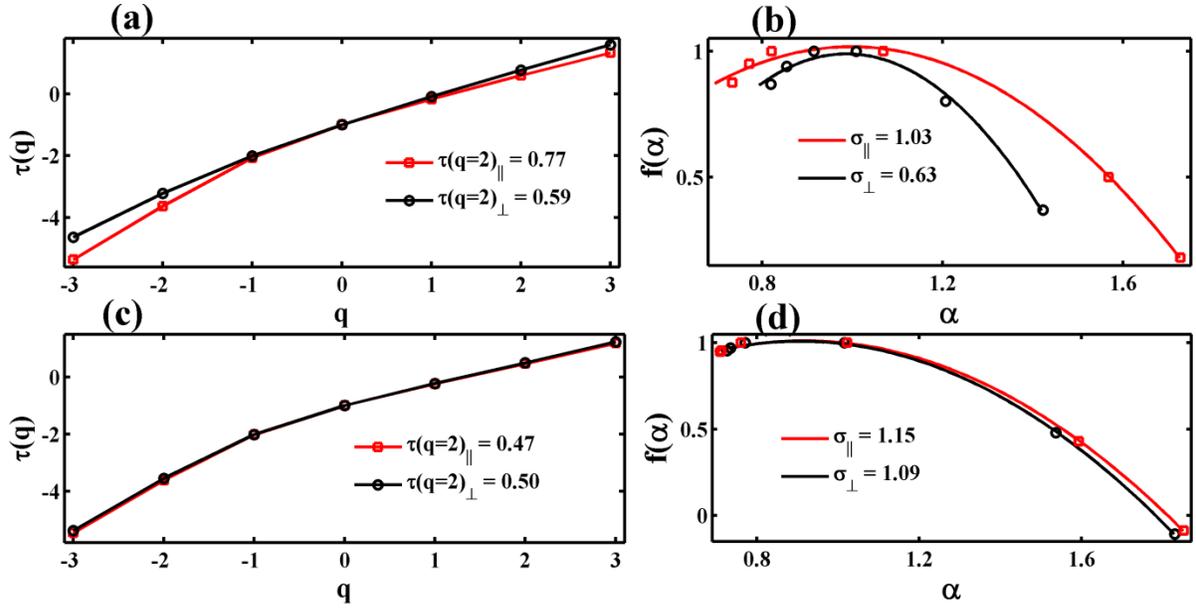

**Supplementary Figure S4:** The results of MFDFA analysis on $\eta_{\parallel}(\rho)$ and $\eta_{\perp}(\rho)$ (derived using inverse analysis [employing Equation (5)] on spectral Mueller matrix elements) from bovine collagen samples before (a and b) and after (c and d) acetic acid treatment. (a) and (c): The moment (q)-dependence of the classical scaling exponent $\tau(q)$; (b) and (d): The corresponding singularity spectra $f(\alpha)$ derived from $\eta_{\parallel}(\rho)$ (red square) and $\eta_{\perp}(\rho)$ (black circle). The values for $\tau(q = 2)$ and width of the singularity spectra ($\sigma$) are noted. The multifractal anisotropy parameters, differential classical scaling exponent, $\Delta\tau = |\tau(q = 2)_{\parallel} - \tau(q = 2)_{\perp}|$; and differential width of singularity spectrum $\Delta\sigma = |\sigma_{\parallel} - \sigma_{\perp}|$ for orthogonal linear polarizations, are observed to decrease following acetic acid treatment. In (a), (b), (c) and (d) lines are guide for eye.